# Toward a new tool to extract the Evidence from a Memory Card of Mobile phones

Rob Witteman, Arjen Meijer
Politie Rotterdam, Team Digitale Opsporing
Doelwater 5, 3011 AH Rotterdam
The Netherlands
{rob.witteman, arjen.meijer}@politie.nl

M-T. Kechadi, Nhien-An Le-Khac
School of Computer Science
University College Dublin
Ireland
{tahar.kechadi,an.lekhac}@ucd.ie

*Abstract*—Today, a mobile phone is not just a "phone" but it is a computer that you can also use for calling someone. Besides, in criminal investigations the importance of evidence from the mobile phone is increasing as more and more phones are seized at the Digital Forensic Department of the police. Indeed, the amount of memory cards of these mobile phones that need to be investigated separately is also increasing. Possible reasons are that the mobile phone investigation software does not support the specific mobile phone or the specific, for that investigation, artefacts. Sometimes the software investigates just the internal memory of the mobile phone and not the data which is written on the memory card. Fact is also that although the mobile phone was investigated by the dedicated software, the possibility that the associated memory card contains additional important information is evident. The current procedure to get all of the usable information from a memory card of a mobile phone is very time-consuming process and not user friendly. In this paper, we present a new single tool to simplify the investigation of a memory card from a mobile phone. We also test our tool with WhatsApp application installed on the memory card from different mobile phones.

## I. Introduction

One of the tasks of a Digital Forensic Investigator is examining a mobile phone. Today a mobile phone is not just for calling, it is more and more expanding towards a little portable computer. The owner of the mobile phone carries a lot of private information in his pocket and in a lot of criminal investigations this private information could be the key to solving the crime. Besides, in the market and literature there are many forensic software that the digital investigators can use in acquisition forensic artefacts from mobile phones. Depending on the brand and type of the mobile phone, relevant information will be extracted. For the digital investigator it is impossible to know for each specific mobile phone which software is the best to investigate.

On the other hand, with the evolution of the mobile phone the device memory was extended from the internal memory to external, replaceable memory. As a consequence, the important information is not only traceable in the device, but also on the memory card. Most of mobile device forensic tools today can retrieve the information from both internal and external memory. However, in some cases, information stored in external memory was not extracted when the mobile phone was examined. Therefore, in order to get a result as complete as possible, it is necessary to examine the memory card separately.

Examining such a little memory card is a challenging task. When examining a memory card, the purpose is to get all the information and deliver it in a friendly readable format to the tactical detective. Why could such a little data object (mostly between 1 and 8 GB) cost so much time to investigate? In fact, this process includes the extraction of all files (pictures, video files, audio files and documents) and recovering delete files but also decrypt chat databases like Whatsapp and put the chat history in a friendly readable format. For the whole process many forensic tools and scripts are normally used. So it is not the amount of data that consumes all the investigation time but the usage of different forensic tools.

Because of the complexity of using different forensic tools and the processing time of memory card investigation, there is a requirement at the Digital Forensic Department of the Police for one forensic tool that automates the whole process.

In this paper, we present our approach to create a new tool that can replace as much as possible different forensic tools required attaining an optimal result when examining a memory card from a mobile phone. We also test our approach with the forensic acquisition and analysis of WhatsApp application installed on the memory card from different models of mobile phones.

The rest of this paper is organised as follows: Section 2 shows related work in this area. We discuss briefly the process of memory card forensics and point out the challenges of this process in Section 3. We present our new tool in Section 4. We describe and analysis results in Section 5. Finally, we conclude and discuss on future work in Section 6.

## II. Related Work

In [1], author describes flash memory forensics. He also presents the architecture of flash memory. He showed moreover hidden data in man-made bad blocks. In fact, His work however focuses on the internal memory rather than external one.

Regan [2] also focuses on the forensic analysis of internal flash memory of mobile phones. He showed how to analyse flash file systems such as YAFFS2 (Yet Another Flash File System used in Android) and the JFFS2 (the Journaling Flash File System). Again, author did not locate at whole investigation process for external memory cards.

Casey et al. [3] described all steps in the process of mobile phone forensics including recovering any deleted items including files, SMS messages, call logs, and multimedia. They also discussed on using different forensic tools such as EnCase [4], FTK [5], BitPim [6] to acquisition and analysis artefacts. This process is widely used in digital forensic department in law enforcement agencies. Authors did not however show how to improve the investigation process.

A research was done in 2011 in the time that "Whatsapp" changed the policy about sending messages [7]. In the first versions of Whatsapp the messages were sent in plain text, but all platforms (IOS, Android, Windows and BlackBerry) changed by encrypting the Whatsapp messages. Therefore the databases stored on the memory card or in the internal memory of the cell phone were also encrypted. The main question in this research was the possibility of decrypting the Whatsapp databases for law enforcement investigation purposes. One of the sub questions was what the differences of encryption between the operating systems IOS, Android and BlackBerry were in.

A research in 2013 [8] on extracting and analysing Whatsapp information from Android cell phones. The area focus was on the volatile memory (RAM) as the non-volatile memory (memory card) of an Android device. The aim of the research is to summarize a general methodology to gather valuable information and developing a standard procedure for all similar applications.

Internet Evidence Finder version 6.3 [9] is commercial software originally intended for analysing internet behaviour, including social network communication as well as chat history of several chat applications. The latest version is provided with a "mobile device" function. It is possible to feed the program with a physical dump of a mobile device or an image file of a storage device such as a memory card. The program is capable to analyse the data and creates a report of predefined artefacts. In the case of Whatsapp history it is not capable to decrypt the BlackBerry chat history.

III. PROBLEM STATEMENT

A. Investigation of a memory card from mobile phones

An investigation of a memory card has normally five steps including the creation of disk image, the extraction files from disk image [3], the file recovery and the conversion/decryption files. In the following paragraphs we briefly describe these steps.

*Create image file*: When the decision was made to examine the memory card, the memory card was ejected from the mobile phone and put in a hardware write blocker [3]. Via the write blocker a connection is made between the memory card and a computer with forensic software "Encase". With Encase a disk image is made of the memory card.

*Extract files from image:* Prior to extracting all the common files from the disk image, a quick visual scan of the contents is performed by the investigator. Depending on file structure, forensic tools can extract relevant files.

*Recover files*: After all interesting files are extracted, investigators use data carving tools such as *"Photorec"* to recover deleted files.

*File decryption*: In some mobile applications, databases stored in memory card are encrypted. In these cases, investigators should use software tools to decrypt databases.

*Report*: When the whole procedure is finished the investigator makes a report of all his findings. This report is a template where all the common findings are already mentioned.

B. Challenges

In a Police Digital Forensic Department of a European country, where a mobile phone is seized, a special detective with a 2-day phone course primarily does the job. If the software (XRY logical, for example) supports the brand and type of the mobile phone and some results are reported. The results of such an extraction of information could be enough for that specific crime, but the first challenge is that not all types of mobile phones are supported and that the supported mobile phones have different outcomes of results. So there still is a need to find more information, further investigation is required at the Forensic Digital Department. Investigators should normally use different forensic tools and it is a time consuming process.

Consequently, there is a need of developing a single tool that simplifies this forensic process. The most important is that the new tool should be simple, automatic and time-saving. Digital investigators should have to complete the procedure without searching all kinds of tools or reading lots of manuals before starting an investigation. Besides, the new tool should be scalable that can easily integrate new libraries to acquire new kind of artefacts. Last but not least, the improvement for new tool is not only a time saving advantage, but is also more user-friendly for the tactical investigators. These requirements are the motivation for us to propose and develop a new tool for the investigation of memory card of mobile devices.

IV. FORMOXTRA TOOL

In this section, we present ForMoXtra, a new single tool that can simplify the investigation of a memory card from a mobile phone. This tool is also able to read a forensic backup of a memory card (image file) and extract all the present files the memory card contains. It can convert files to a common format if necessary. ForMoXtra is composed of the following components: file extraction, file recovering and analyzing, file converting and decrypting, reporting and GUI (graphic user interface) (Figure 1).

*File extraction*: the input of ForMoXtra is the disk image files created from memory cards. Prior to the extraction, a quick scan of the contents is carried out to identify the mobile phone operating system. For each type of mobile phone operating system, relevant file categories will be extracted. In most cases, they are pictures, video-files, audio-files and documents. For example, in the case of BlackBerry, the file categories are pictures, video-files, audio-files, thumb.dat, key.dat and documents.

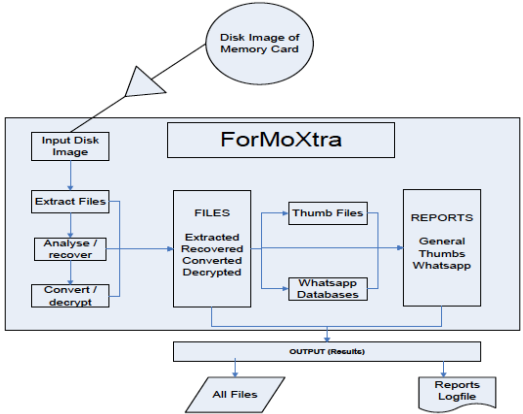

Fig. 1. ForMoXtra architecture and process flow.

*File recovering/analysing*: After all interesting files are extracted, a procedure to recover deleted files is started. The recovering module uses data carving methods such as FileFinder tools (Encase). All the extracted files and the recovered files are sorted in categorized folders. In this module we also integrate "Photorec" tool as it is capable of recovering deleted files directly from a disk image. The recovered files should be placed in predefined (recovery) folders.

*File converting and decrypting*: This component is responsible for converting all non-common audio and video files should be converted to .wma and .wmv format. The original format should always remain and is also part of the output. The thumb files, containing pictures and additional information, will be also processed into a readable report style output. This component moreover decrypts popular encrypted databases such as Whatsapp databases into a readable format. The contents of the current and the backup databases are also compared and combined in one report.

*Reporting and GUI*: This component offers GUI interface to the investigator that he can enter relevant information of the case including Case number, Object number, location of the disk image and the location of the output, etc. (Figure 2).

## V. PERFORMANCE EVALUATION AND ANALYSIS

In this section, we evaluate and analyse ForMoXtra tool. In fact, it is hard to build one tool that can cover all steps of a forensic process mentioned in Section III. This ForMoXtra tool can actually simplify this process. So creating disk images of memory cards is not a part of this tool. We use EnCase in our experiments to take the disk image and store in a working folder and then we start our FoxMoXtra tool. In this section, we test ForMoXtra tool by analysing memory cards stored artefacts from WhatsApp application on mobile devices.

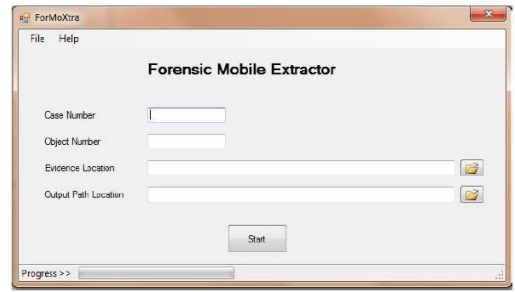

Fig. 2. Input screen ForMoXtra.

### A. Experiments

Whatsapp changed main policies in the Android version considering storage and encryption. The first change was that the Whatsapp databases can only be stored in device memory and not on the memory card. The second main change was that for decryption not only was the known encryption key required but at least also the account email address of the mobile phone which the memory card belongs to. The decryption of Android Whatsapp is integrated in ForMoXtra but only for older Whatsapp versions. Since Android version 2.11.242 / 2.11.244 (March, 2014) it is not possible to decrypt without knowing the account name of the Whatsapp user. The filenames of these Whatsapp databases are recognizable by the addition "crypt5" or "crypt7". It is not possible anymore to use Whatsapp for messaging with an older version, so updating to this new version is required. The consequence is that all future investigations of the Whatsapp history on memory cards from an Android mobile phone, cannot be done when the account name is not known. More research has to be done to find out which procedure has to be followed to decrypt those databases. Therefore the decrypting of Android Whatsapp is not part of this paper.

For BlackBerry Whatsapp databases the procedure is still the same. Every BlackBerry mobile phone creates its own encryption key which it stores on the device. For obtaining that key, access to the mobile phone is still required. When the mobile phone is locked, the user code is required to get access. Once the mobile phone is unlocked, the described tool "BlackBerry Desktop Software" is used to create a ".bbb" file. So, in case of a BlackBerry Whatsapp database, the input for ForMoXtra should be the disk image and the ".bbb" file. Briefly, in our experiments, we use Encase and BlackBerry Desktop to create the input data for our ForMoXtra tool.

In terms of materials, in our experiments, we use two mobile phones: Blackberry 9300 (test name: 93, RIM, 2GB SD), Blackberry 9800 (test name: 98, RIM, 2GB SD).

### B. Analyse

After installing Whatsapp, a SQLite database with the filename "messagestore.db" was created on the memory card. On a BlackBerry 9300 it is not possible to store Whatsapp history in the device memory. During the testing period that database with the name "messagestore.db" did not change physical location on the memory card. When messages and pictures were sent or deleted, the contents and hash value were changed. Size changed not with every message but when only when the reserved database space was exceeded. This happened by sending some pictures.

When we update information in the database, a backup Whatsapp SQLite database was created and stored at a free location on the memory card. Such a backup file is named like "messageStore.db.w.1392197250043.bak", where the number represents a Unix epoch time of the creation time of the backup file. This backup file has the same hash value as the previous "messagestore.db" file. Even when nothing was changed, when no Whatsapp conversation had taken place, nevertheless a backup file was created. It got the same hash value as the current "messagestore.db" file. Those creations of backup files occurred at different moments, for example after disconnecting from power or re-inserting the memory card.

In the test, a change was made to the current "messagestore.db" without sending messages. The application "Whatsapp" was started and we found that with a BlackBerry 9300 it was not possible to delete a single message. No further action was taken, but it was enough to change the database "messagestore.db". Two new database files were also created. Because of the Timestamps of "Last Written" it seems that the "messagestore.db" was changed two times between the tests: first after sending a message and the second after reading all incoming messages.

During our experiments, bak files were created but also deleted. The total number of SQLite databases grew, including the current "messagestore.db" to 7 databases. The content of the very first "messagestore.db" database was still present at the end of testing in the backup database with the filename "messageStore.db.w.1392197250043.bak. After these tests, there were 6 backup "Whatsapp" databases and one current "Whatsapp" database.

Conclusion on these testing is that changes (adding or deleting messages) made in the Whatsapp messaging are stored in the current database and in the backup database created after the changes were made. The previously created backup database does not contain these changes, so new messages and deleted messages are present in that backup database as long as this complete database is not deleted and overwritten.

The message table of a BlackBerry Whatsapp database contains 28 fields. However, not all the fields are interesting for crime investigation so to make it readable only the following fields are of interest and therefore present in the output. (Figure 3)

To search for unique messages, ForMoXtra starts examining the oldest backup Whatsapp database and extract all messages. All messages are sorted by Contact name (key_remote_jid). In order to indicate to which contact the message history belongs, this Contact name is placed above the messages. Each message contains the fields "timestamp", "key_from_me" and "data" sorted by timestamp. To each message a field with the filename of the examined database is added.

Next, the second oldest backup Whatsapp database is examined. Based on a unique combination of table fields (key_remote_jid and timestamp) it is checked if messages already exist in the previous examined database. If a message was also present in the previous database, the filename changes in the filename of the second oldest database. This procedure goes on until the examination all the available databases is finished.

| Field | Description |
|---|---|
| key_remote_jid | (31612345678@s.whatsapp.net) the part before @s.whatsapp.net represents the phone number of the chat contact |
| key_from_me | Has values "0" = incoming or "1" = outgoing |
| Data | Message content |
| Timestamp | Timestamp is in Unix UTC format |

Fig. 3. Important fields of table "messages" of a BlackBerry Whatsapp database.

Next, we create a HTML document where the variable input fields as: Casename, Casenumber and Itemnumber are presented. Each contact is presented by their phone number. The phonenumbers are linked to the associated chat history. The chat history is sorted by contact and sorted by date/time. If the message contains a picture, a thumbnail of the picture with the note "picture" is presented. When the picture is still present at the memory card, this note "picture" is also hyperlinked to the picture stored at the memory card. The column "Filename" contains the name of the decrypted "Whatsapp" database that the message belongs to. Figure 4 is shown that all the messages are present in the current "Whatsapp" database (messageStore.db_decr) except the message with timestamp "2014-02-24 14:15:14" in chatsession "316213…...". This message belongs to a backup database with filename "messagestore.db.d.1393504046326.bak", which indicates that that message was deleted. Figure 4 also represents the HTML output of "Whatsapp" as a result of the "ForMoXtra" procedure.

The behaviour of the BlackBerry 9800 Whatsapp databases appeared to be the same as the BlackBerry 9300. Once the databases are decrypted they act like SQL databases and have the same database structure as those from the BlackBerry 9300.

*Compare test results with other tools:* The testing period was a period that Android Whatsapp decrypting was possible with only the known encryption key. Very few commercial software and open source tools have been developed to extract Whatsapp from mobile phones and memory cards. To discover the advantages and disadvantages of these tools we tested with 3 common commercially available programs and the open source tool of Zena Forensics [10]. The tools XRY [11] and UFed 12] are developed for investigating mobile phones with or without a memory card but not for memory cards on their own. IEF (Internet Evidence Finder) is a commercial tool. The standard version is capable of reading disk images of memory cards and also decrypting Android Whatsapp databases (older versions) and analyse them but not for BlackBerry. The advanced and more expensive version is also capable of analysing physical dumps of a mobile phone.

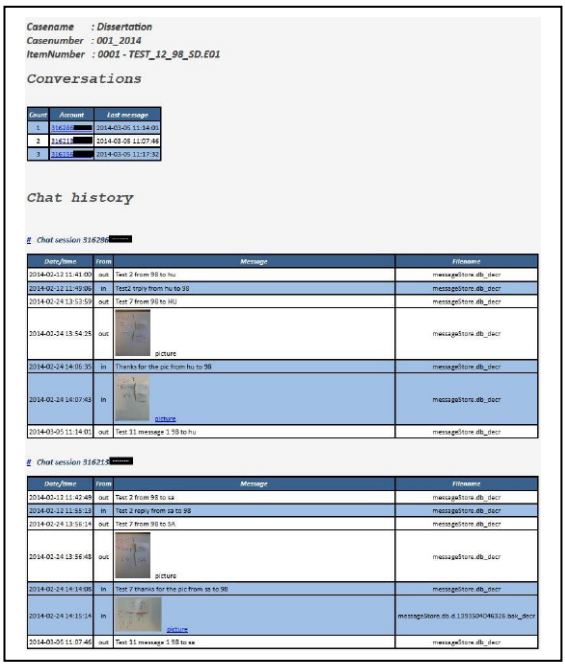

Fig. 4. Output of combined Whatsapp databases.

*Further experiments:* We also tested 6 different memory cards from different BlackBerry mobile phones. The scope of comparison was: the extraction of the files, the decryption and presentation of Whatsapp. Besides accuracy of automated extraction, time savings could be a big success factor.

The average time ForMoXtra uses to come with the desired result is 15 minutes there were the average manual time was 45 minutes. In this time is included the creation of the evidence file and the BlackBerry Backup file. These two steps are in both procedures the same.

Besides the time saving with the "ForMoXtra" tool, the simplicity has improved a lot. In the manual procedure three tools were required for presenting "Whatsapp" history. Encase for extracting the databases, Elcomsoft BlackBerry [13] for decrypting the databases and the python script for putting the databases in a user friendly output. This is now simplified to one tool: ForMoXtra. In the case of processing a large amount of forensic data from memory card, we are also looking at a knowledge map solution [14] to interact with "ForMoXtra" tool to handle the forensic results more efficient.

## VI. CONCLUSION AND FUTURE WORK

In this paper we present a new single tool to simplify the investigation of a memory card from a mobile phone. We also test our tool with WhatsApp application on different mobile phones. Through our experiments, this tool is much simpler as manual process. The improvement of the output was very welcome for the tactical detectives in our Digital Department. Although the time saving is significant already, the integration of the other features to make this tool more powerful is considered. Besides that, when ForMoXtra is finished as proposed, it could be of great value for the specialist in the rest of the police region. We are also looking at the feasibility of using this tool in forensic acquisition of Instant Message Applications [15].

REFERENCES

[1] S. Fiorillo, "Theory and practice of flash memory mobile forensics", Proceedings of the 7th Australian Digital Forensics Conference, December 2009, Perth, Western Australia

[2] J. E. Regan, "The forensic potential of flash memory", MSc Thesis, Naval Postgraduate School, September 2009, Monterey, California, USA

[3] E. Casey, "Digital Evidence on Mobile Devices", Chapter 20 in Digital Evidence and Computer Crime: Forensic Science, Computers, and the Internet, 3rd Edition, Elservier Publisher, November 2011

[4] EnCase Forensics https://www2.guidancesoftware.com/products/Pages/encase-forensic/overview.aspx

[5] Forensic ToolKit http://accessdata.com/solutions/digital-forensics/forensic-toolkit-ftk

[6] BitPim http://www.bitpim.org/

[7] D. Cortjens, A. Spruyt, and W. F. C. Wieringa - "WhatsApp Database Encryption Project Report."-2012 https://www.os3.nl/_media/2011-2012/courses/ssn/whatsapp_database_encription_on_android_and_BlackBerry.pdf.

[8] Neha S. Thakur – "Forensic Analysis of WhatsApp on Android Smartphones" – New Orleans – 2013 - http://josemilagre.com.br/blog/wp-content/uploads/2014/03/Forensic-Analysis-of-WhatsApp-on-Android-Smartphones.pdf

[9] Internet Evidence Finder by Magnet Forensics Inc. - http://www.magnetforensics.com/software/internet-evidence-finder/ief-advanced/

[10] Zena Forensics http://blog.digital-forensics.it/2012/05/whatsapp-forensics.html

[11] XRY XRY MicroSystemation A.B. (MSAB) - http://www.msab.com/xry/what-is-xry

[12] UFED Touch Ultimate, Cellebrite, - http://www.cellebrite.com/mobile-forensic-products/ufed-touch-ultimate.html

[13] Elcomsoft BlackBerry Backup Explorer - http://www.elcomsoft.com/ebbe.html

[14] N-A. Le-Khac, L. Aouad and M-T. Kechadi – "Distributed knowledge map for mining data on grid platforms", International Journal of Computer Science and Network Security, Vol.7(10), pp.98-105

[15] C. Sgaras, M-T. Kechadi and N-A. Le-Khac – "Forensics Acquisition and Analysis of Instant Messaging and VoIP Applications", Computational Forensics, Springer Verlag, Lecture Notes in Computer Science, Vol. 8915, p.188-199, 2015